\documentclass[12pt]{article}
\input{epsf.sty}
\usepackage{hyperref}
\usepackage[pdftex]{graphicx}
\usepackage{amsmath, amssymb}
\newtheorem {thm}{Theorem}[section]
\newtheorem {prop}[thm]{Proposition}

\newtheorem {cor}[thm]{Corollary}
\newtheorem {defn}[thm]{Definition}

\def\Cox{\hfill \Box}

\def\N{{\Bbb N}}

\def\Z{{\Bbb Z}}
\def\R{{\Bbb R}}

\def\P{{\Bbb P}}

\def\e{{\varepsilon}}

\def\S{{\Bbb S}}

\def\eps{\varepsilon}

\def\ba{{\backslash}}

\def\D{\Delta}
\def\a{\alpha}

\def\ba{\setminus}
\def\b{\beta}
\def\MM{{\cal M}}
\def\d{\delta}

\def\e{\varepsilon}

\def\phi{\varphi}

\def\g{\gamma}

\def\l{\lambda}

\def\s{\sigma}

\def\z{\zeta}

\def\o{\omega}

\def\D{\Delta}

\def\L{\Lambda}

\def\G{\Gamma}

\def\O{{\Omega}}

\def\P{{\Phi}}

\def\T{\T}

\def\PP{{\cal P}}

\newcommand{\triplenorm}{| \hspace{-0.3mm} | \hspace{-0.3mm} |}

\begin{document}
\title{Gibbs-non-Gibbs properties for n-vector lattice and mean-field models}

\author{
Aernout C.D. van Enter
\footnote{ University of Groningen, Institute of Mathematics and
Computing Science, Postbus 407, 9700 AK Groningen, The
Netherlands,
\newline
 \texttt{A.C.D.van.Enter@rug.nl}, \texttt{
http://statmeca.fmns.rug.nl/ }}
 \, , Christof K\"ulske
\footnote{ 
\texttt{c.kulske@rug.nl}, \texttt{
http://www.math.rug.nl/$\sim$kuelske/ }}\, 
, Alex A. Opoku 
\footnote{ \texttt{A.a.opoku@math.rug.nl} \texttt{
http://statmeca.fmns.rug.nl/Alex.html }}
\, \\ and
Wioletta M. Ruszel
\footnote{ \texttt{W.M.Ruszel@rug.nl} \texttt{
http://statmeca.fmns.rug.nl/wruszel.html }}
}

\maketitle

\begin{abstract} 
We review some recent developments in the study of Gibbs and non-Gibbs properties 
of transformed $n-$vector lattice and mean-field models under various transformations. Also, some new results for 
the loss and recovery of the Gibbs property of planar rotor models during 
stochastic time evolution are presented.

 \end{abstract}

\smallskip
\noindent {\bf AMS 2000 subject classification:} 82B20,
82B26, 60K35.

 \smallskip
\noindent {\bf Keywords:} Gibbs measures, non-Gibbsian measures, $n-$vector lattice models, $n-$vector mean-field models, 
 transformed model, Dobrushin uniqueness, cluster expansion, spin-flop transitions.

\vfill\eject

\section{Introduction}

In the recent decade and a half there has been a lot of activity on the topic
of non-Gibbsian measures. Most of the original studies were 
based on the question of whether renormalised Hamiltonians exist as properly 
defined objects, 
\cite{ACDR,GP78,GP79}, with an emphasis on  discrete-spin models. 
Another issue, which also arose  in physics but somewhat later \cite{OP}, 
was the following question:  
Apply a (stochastic) dynamics which converges to a system at a temperature $T_1$ to an 
initial state at temperature $T_2$ for a finite time. Is the resulting
measure in this transient non-equilibrium regime a Gibbs measure?  
Could it be described in terms of an effective temperature (hopefully between initial and final one)?  
Again the first results \cite{ACD1} were for discrete spins.
Afterwards more general dynamics and also unbounded spins were investigated in 
\cite{LEN,ROE,KUL5}. Although the work of \cite{ROE,KUL5} was about 
continuous spins, there remains something of a problem, in that for unbounded 
spins the notion of what one should call Gibbsianness for a ``reasonable'' 
interaction is less clear than in the compact case. Thus it turned out to 
be of interest to see 
how a model with compact but continuous spins behaves. Another extension of 
the original investigations was the investigation into the question of what the proper mean-field 
version of the Gibbs-non-Gibbs question might be. For this, see in particular
\cite{KUL1,KUL2,HaggKu04}. This question has a particular charm  
for systems with a general local spin space. 

As there have recently been a number of reviews on other aspects of the non-Gibbsian problem \cite{LEN1,LEN2,AFE08,DEZ,ACD}, we here want to emphasize 
what has been found for $n$-vector spins. The results as yet are less complete 
than what is known for Ising or Potts spins, but it has also become clear that,
although many things are similar, such systems have traits of their own 
which are 
somewhat different and require new ideas. We have mostly worked on 
transformations such as stochastic evolution, which does not rescale space, 
such as renormalisation group transformations do. Note that in a statistical 
interpretation, such maps for discrete spins model imperfect observations, 
that is observations in which with some probability one makes a mistake, an 
interpretation which already was mentioned in \cite{GP79}. For continuous spins, the probability of staying exactly at the initial value is zero, but for 
short times the map is close to the identity in the sense that 
the distribution of an evolved spin is concentrated on a set close to 
the initial value. 
We obtained conservation of Gibbsianness under stochastic evolutions when 
either the time is short, or when both initial and 
final temperature are high. We also found that loss of Gibbsianness occurs
if the initial temperature is low, and the dynamics is an infinite-temperature 
one. If the initial system is in an external field, after a long time the measure can become Gibbsian again. In fact, here we extend the regime where such results can be proven.

Another question we could address is the discretisation question. If one approximates a continuous model by a discrete one, is the approximation still a Gibbs measure, now for discrete spins? 
Morally, this question is somewhat related to renormalisation-type questions, as in both cases some coarse-graining takes place, 
in which the transformed system only contains part of the initial information. 
It turns out that the transformed measure is Gibbsian, once 
the discretisation is fine enough. All of these questions, the 
high-temperature and short-time Gibbsianness for stochastic evolutions, as well as the loss and recovery properties,  
can also be addressed in the mean-field setting, and we find that the results are similar as in the lattice case. 
Again, for transformations which in some sense are close enough to the identity, the transformed model is Gibbsian.
Finally, one may ask which of our results depend on the fact that our local state space is a sphere and not 
just a compact space? The regularity results (preservation of Gibbsianness) do not, 
as they are based on absence of phase transitions. 
In fact, such extensions have been proved, for which we refer to the original papers. 
When it comes to a failure of Gibbsianness, an internal phase 
transition has to be exhibited. The mechanism of this is usually very model-dependent 
and this is where the intricacy but also the charm of the $n$-vector models lies. 

\section{Gibbsianness and non-Gibbsianness for $n$-vector lattice models}
In this section we review some recent developments in the study of Gibbsianness and non-Gibbsianness for 
$n-$vector models subjected to various transformations. The review is mainly  based on the recent papers \cite{WIO,WIO1,KULOP}. 
Before we plunge into details let us fix some definitions, notation and give some 
background from the theory of lattice spin systems. 

\subsection{Notation and Definitions}
For general information on Gibbs measures for lattice spins systems we refer the reader to \cite{ACDR,GOR}.
In this review we will focus attention on models living on a $d-$dimensional lattice $\Z^d$ ($d\geq1$).
We will take $\S^n$, the $n-$dimensional sphere, as the single-site spin space equipped with a Borel probability 
measure $\a$ (the a priori measure). The measures we will study shall be given 
by Hamiltonians. The Hamiltonians in a finite volume $\L\subset \Z^d$, with boundary condition $\o$
outside $\L$, will be given by
\begin{equation}\label{hamil}
H_\L^{\o}(\s)=\sum_{A:\  A\cap \L\neq \emptyset}\P_A(\s_\L\o_{\L^c}),
\end{equation}
where the interaction $\P$ is a translation-invariant family of functions $\P_A:(\S^n)^{\Z^d}\rightarrow \R$, 
with $\P_A$ depending only on the spins in the finite volume $A$. 
It satisfies the following absolute summability condition

\begin{equation}\label{absum}
\|\P\|=\sum_{ A \ni 0}\|\P_A\|_\infty<\infty.
\end{equation}

The \textbf{Gibbs measures } for the interaction $\P$ are the measures $\mu$ on $(\S^n)^{\Z^d}$ whose finite-volume 
conditional distributions are given by
\begin{equation}\label{Gibbs}
\mu_\L(d\s_\L|\o_{\L^c})=\frac{\exp\left(-H_\L^{\o}(\s)\right)\a^\L(d\s_{\L})}{Z_\L^\o},
\end{equation}
where $\a^\L$ is the product measure of $\a$ over the sites in $\L$. Another, equivalent, way 
of defining a Gibbs measure was identified by Kozlov \cite{Koz}, 
via two properties of the family of conditional distributions $(\mu_\L)_{\L}$. These 
properties are \textbf{uniform nonnullness }and \textbf{quasilocality}. The latter property holds for a measure $\mu$
 if for all $\e>0$, $i\in\Z^d$ and configurations $\eta$ there exists a $\L\ni i$ and $\G\supset \L$ such that for all 
 pairs of configurations $\o,\zeta $
 
 \begin{equation}\label{quasiloc}
\big|\mu_\G(\s_i|\eta_{\L\ba i}\o_{\G\ba\L})-\mu_\G(\s_i|\eta_{\L\ba i}\zeta_{\G\ba\L})\big|<\e.
\end{equation} 
A collection $\g$ of everywhere defined conditional distributions $\g_\L=\mu_\L$ satisfying all the above conditions is referred to as 
a \textbf{Gibbsian specification}.

Now what can be said about the Gibbs properties of transformed Gibbsian 
$n-$vector models? In \cite{WIO,WIO1,KULOP}
the Gibbs properties of various transformations acting on $n-$vector models 
were investigated and we will review the results below.


\subsection{Conservation of Gibbsianness under local transformations close to the identity}
We discuss conservation of Gibbsianness for initial Gibbsian $n-$vector lattice 
models subjected to local transformations close to the identity. The discussion will mainly 
follow \cite{WIO,KULOP}. Though these two papers use different techniques, the results 
proved therein have some common ground and we will compare
the advantages and disadvantages of both methods. We will mainly address conservation of 
Gibbsianness for transformed initial Gibbs measures in this subsection.

We start with a Gibbs measure $\mu$ of an $n-$vector model and apply local transformations to it. Examples of such 
local transformations are infinite-temperature diffusive dynamics (sitewise independent Brownian motions on spheres), 
fuzzification or discretisation of the local spin space, etc. 
The natural question that comes to mind is whether such a transformed measure
$\mu'$ is a Gibbs measure. For transformations close to identity the above question can be answered in the affirmative. This we make
precise in the sequel by first stating a theorem which is the intersection of the results found in \cite{WIO,KULOP}. 
 
\begin{thm}\label{mainthm}
Suppose $\mu$ is the Gibbs measure for a translation invariant interaction $\P$ on $(\S^1)^{\Z^d}$. Further, assume 
that $\P$ is twice continuously differentiable and of finite range. Let $\mu_t$ be the transformed (time-evolved) 
measure obtained by applying 
infinite-temperature diffusive dynamics to $\mu$. Then for short times the time-evolved measure $\mu_t$ is a Gibbs measure.
\end{thm}
Theorem \ref{mainthm} can be proved either by using cluster expansion techniques as 
in \cite{WIO} or by Dobrushin uniqueness techniques 
\cite{KULOP}. The results proved in these papers generalise the above theorem
  in different directions. In the following we will 
review some of the main issues discussed in them.
Let us start with the approach of \cite{WIO}. 
The advantage of using cluster expansion techniques is that we can prove short-time 
Gibbsianness for more general dynamics beyond the independent Brownian motion on the circles. 
In particular, one can handle  a whole class of systems 
which are modeled via the solution $\sigma=(\sigma_i)_{i \in \Z^d}(t)$ of the 
following system of {\em interacting} stochastic differential equations: 
\begin{eqnarray}
\begin{cases}
& d \sigma_i(t) = -\nabla_i \frac{1}{2} \beta_1 H_i^{d}(\sigma(t))dt + d B_i^{\odot}(t) , t > 0, i \in \mathbb{Z}^d \\
& \sigma(0) \simeq \mu , t=0 \label{system}
\end{cases}
\end{eqnarray} 
where $(B_i^{\odot}(t))_{i,t>0}$ denotes a family of independent Brownian motions moving on a circle, $\nabla_i=\frac{d}{d\sigma_i}$ and $\beta_1 \sim 1/T_1$ is the ``dynamical'' inverse temperature. We assumed that the ``dynamical'' Hamiltonian $H^{d}$ is built from an absolute summable ``dynamical'' interaction which is again of finite range and at least twice continuously differentiable. Let $S(t)$ denote the semigroup of the dynamics defined in \eqref{system}. Then one can prove that 
for all values of $\beta_1$ the time-evolved measure $\mu_t=\mu \circ S(t)$ is Gibbsian for 
short times.
Note that the statement of Theorem \ref{mainthm}
corresponds to the case where $\beta_1=0$. We note that the cluster expansion technology was heavily influenced by \cite{ROE}. Extensions to different graphs are also immediate. \\

The proof in \cite{WIO} makes also use of the fact that $\mathbb{S}^1 \simeq [0,2\pi)$ where $0$ and $2\pi$ are considered to be the same points. Consequently we can work on the real line and do not have to worry about more general compact manifolds $\mathbb{S}^n$. Although it is in principle possible to write a cluster expansion for $\mathbb{S}^n$ and we believe that short-time Gibbsianness for general interacting dynamics holds also in higher spin dimensions, this has not been done so far.

\bigskip
 
Next let us review the results in \cite{KULOP}. The Dobrushin uniqueness
technique employed in that work applies to more general interactions 
on general $n$-spheres and also to more general graphs aside from $\Z^d$. 
Moreover, we expect that it provides better bounds for the Gibbsian regime 
than the cluster expansion approach does. On the other hand, no results for 
interacting dynamics have been obtained via this approach, 
although we believe in principle this should be possible. 

To be precise, 
one considers initial Gibbs measures for interactions $\P$ with finite triple norm, i.e.
\begin{equation}\label{triple}
\triplenorm \P\triplenorm :=\sup_{i\in \Z^d}\sum_{A: \;A\ni i}|A|\Vert\Phi_{A}\Vert_{\infty}<\infty.
\end{equation}
Note that
this summability condition implies the one in \eqref{absum} and here we don't require that $\P$ is translation invariant.
Initial Gibbs measures of such interactions were 
subjected to local (one-site) transformations given by $K(d\s_i,d\eta_i)=k(\s_i,\eta_i)\a(d\s_i)
\a'(d\eta_i)$, with $\|\log k\|_\infty<\infty$. Here $\eta$ represents the spin variable for the transformed 
system taking values in a compact separable metrizable space $S'$, which now 
needs not be the same as $\S^n$.

In the language of Renormalization Group Transformations, one could think of the transformed system as the 
renormalized system obtained via the single-site renormalization map $k$. The map $k$ can also be thought of as the 
transition kernel for an infinite-temperature dynamics, where the variable in the second slot will be the 
configuration  of the system at some time after starting the dynamics from the 
configuration  in the first slot of $k$. Sometimes we will refer to the time direction as the ``vertical'' direction.

Starting with an initial Gibbs measure $\mu$ for an interaction $\P$ with finite triple norm, 
in \cite{KULOP} it was studied to what extent the transformed measure 
\begin{equation*}\label{eq:aim}
\begin{split}
&\mu'(d\eta):=\int_{\O}\mu(d\s)\prod_{i\in \Z^d}K(d\eta_i|\s_i).
\end{split}
\end{equation*}
will be Gibbsian. In the above we have set $\O=(\S^n)^{\Z^d}$. The study in \cite{KULOP} 
uses Dobrushin uniqueness techniques. The paper also provides continuity estimates for the single-site conditional distributions 
of the transformed system whenever it is Gibbsian. To introduce these estimates the authors 
made use of a so-called 
``goodness matrix'', which describes the spatial decay of the conditional distributions of the transformed measure. \\
In the sequel we will write $i$ for $\{i\}$ and $i^c$ for $\Z^d\ba \{i\}$. In particular the following definition from \cite{KULOP} will be used.

\begin{defn}\label{otto} 
Assume that $d$ is a metric on $\S^n$ and $Q=(Q_{i,j})_{i,j\in \Z^d}$ is a non-negative matrix with 
$\sup_{i\in \Z^d}\sum_{j\in \Z^d}Q_{i,j}=\Vert Q\Vert_{\infty} < \infty$. 
A Gibbsian specification $\g$ is said to be of goodness $(Q,d)$ if the single-site parts $\g_i$ satisfy the 
continuity estimates 
\begin{equation}\begin{split}\label{deno-1}
&\Big\Vert\g_i(d\eta_{i}|\eta_{i^c})-\g_i(d\eta_{i}|\bar\eta_{i^c})\Big\Vert
\leq \sum_{j\in i^c} Q_{i, j }d(\eta_j,\bar\eta_j).
\end{split}
\end{equation}
Here $\Vert \nu_1-\nu_2\Vert$ is the variational distance between the measures $\nu_1$ and $\nu_2$.
\end{defn} 
The matrix $Q$controls the influnce on the specification due to variations 
in the conditioning when we measure them in the metric $d$. The faster $Q$ 
decays, the better, or ``more Gibbsian'', the system of conditional 
probabilities is. We note, without going into details, that a fast decay of $Q$
also implies th existence of a fast decaying interaction potential, but in our 
view an estimate of the form (7) is more fundamental than a corresponding 
estimate on the potential.

We are restricting our attention to single-site $\g_{i}$'s since all $\g_{\L}$ for finite $\L$ can be 
expressed by an explicit formula in terms of the $\g_i$'s with $i\in\L$. For the solution of this 
"reconstruction-problem" see \cite{GOR,FeMa06}. 

The Dobrushin interdependence matrix $C=(C_{ij})_{i,j\in \Z^d}$ of a Gibbsian specification $\g$ 
\cite{GOR,DOB}, is the matrix with smallest matrix-elements for which the specification $\g$ is of the 
goodness $(C,d)$. Here $d$ is the discrete metric on $\S^n$ given by $d(\eta_j,\eta_j')=
1_{\eta_j\neq \eta'_j}$. 

One says that $\g$ satisfies the \textbf{Dobrushin uniqueness condition} whenever the Dobrushin constant $\sup_{i\in \Z^d}\sum_{j\in
 i^c}C_{ij}<1$, and such a Gibbsian specification $\g$ admits a unique Gibbs measure \cite{DOB,GOR}.

Let us now introduce some notation for our discussion on conservation of Gibbsianness for transforms of Gibbs measures.
Set for each $i\in \Z^d$, $\a_{\eta_i}(d\s_i):=K(d\s_i|\eta_i)$, the a priori measures
on the initial spin space which are obtained by conditioning on transformed spin configurations.
We call
\begin{equation}\label{naturalmetric}
\begin{split}
&d'(\eta_i,\eta'_i):=\Vert \a_{\eta_i}- \a_{ \eta'_i}\Vert
\end{split}
\end{equation}
the posterior (pseudo-)metric, associated to $K$ on the transformed spin space $S'$.
$d'$ satisfies non-negativity and the triangular inequality, but we may have $d'(\eta_i,\eta'_i)=0$ 
for $\eta_i\neq \eta_i'$ (which happens e.g. if $\s_i$ and $\eta_i$ are independent under $K$).
For any given $\P$ with finite triple norm write ${\rm std}_{i,j}(\P)$ for 
\begin{eqnarray}\label{stdphi}
{\rm std}_{i,j}(\P):=\sup_{\eta_i\in S'}\sup_{{\z,\bar\z\in \O}\atop{
\z_{j^c}=\bar\z_{j^c}}} \inf_{b\in\R}\Bigg(\int \a_{\eta_i}(d\s_i)\Big(
H_i^{\z}(\s_i)-H_i^{\bar\z}(\s_i)-b\Big)^2\Bigg)^{\frac{1}{2}},
\end{eqnarray}
where $H_i^{\bar\z}(\s_i)$ is as in \eqref{hamil}.

Consider the matrix
\begin{equation}\begin{split} 
& \bar C_{ij} :=\frac{1}{2}\exp\Bigr(\sum_{A\supset\{i,j\}}\frac{\delta(\Phi_A)}{2}\Bigl){\rm std}_{i,j}(\P).
\end{split}
\end{equation} 
Here we have denoted by $\d(f)$, for $f$ a real-valued observable on $\O$, the oscillation of $f$ given by
$$\d(f):=\sup_{\o\neq \xi}|f(\o)-f(\xi)|.$$ 
The above quantity $\bar C_{ij}$, can be small if either  the initial interaction $\P$ is weak or  the measures $\a_{\eta_i}$ 
are close to  delta measures. For example this is the case for short-time evolution of the initial Gibbs measure
 associated with $\P$, as we will point out later. 
$\bar{C}$ is an upper bound 
on the Dobrushin matrices for the joint systems consisting of the initial and the transformed spins vertically coupled 
via the map $k$, and having fixed transformed configurations. 
A specification for this sytem is generated by $\Phi$ by replacing in equation 
(3) the a priori measure $a$ by the $a_{\eta_i}$'s.
The main tool used in \cite{KULOP} to 
show Gibbsianness of the transformed measure was the lack of phase transitions in the conditional joint system discussed 
above. This lack of phase transitions will follow if the Dobrushin constant of the matrix, $\bar C$, is strictly less than $1$. 
More precisely the following theorem was proved (\cite{KULOP}: Theorem 2.5). 
\begin{thm}\label{mainthm-lat} 
Suppose that $\mu$ is a Gibbs measure associated with a lattice interaction $\P$ with finite triple norm. 
Suppose further that $\sup_{i\in \Z^d}\sum_{j\in i^c}\bar C_{i,j}<1$. Then the transformed measure 
$\mu'$ is Gibbsian and the transformed Gibbsian specification $\g'$ has goodness $(Q,d')$, where

\begin{equation}\begin{split}\label{dino}
& Q_{i j }= 4 \exp\Bigl(4\sup_{i\in\Z^d} \sum_{A\ni i}\Vert \P_A\Vert_\infty\Bigr)
\sum_{k\in i^c}\d_k\Bigl( \sum_{A\supset \{i,k\}}\Phi_{A} \Bigr)
\bar D_{k j}
\end{split}
\end{equation}
with $\bar D=\sum_{n=0}^\infty \bar C^n$.
\end{thm}
Thus the transformed measure $\mu'$ will be Gibbsian if either the initial interaction $\P$ is weak
or the a priori measures $\a_{\eta_i}$ are close to delta measures. Furthermore, in the above theorem the goodness of 
the transformed specification is expressed in terms of the posterior metric $d'$. Can one have local 
transformations where this metric could be expressed in terms of more familiar metrics on $\S^n$? In what 
follows we present two examples where the above question will have a positive answer. To do 
this we pay a price of putting further restrictions on the class of allowed interactions for the initial system.

\begin{defn}
Let us equip $\S^n$ with a metric $d$. 
Denote by $L_{ij}=L_{ij}(\Phi)$ the smallest constants such that the
$j$-variation of the Hamiltonian $H_i$ 
satisfies 
\begin{equation}\begin{split}\label{lippi}
& \sup_{{\zeta,\bar \zeta}\atop{\zeta_{j^c}=\bar \zeta_{j^c}}} \Bigl
| H_i^{\z}(\s_i)-H_i^{\bar\z}(\s_i)-\Big(H_i^{\z}(a_i)-
H_i^{\bar\z}(a_i)\Big)\Bigr |\leq L_{i j }d(\s_i,a_i).
\end{split}
\end{equation}
We say that $\Phi$ satisfies a Lipschitz property with constants $(L_{ij}(\Phi))_{i,j\in \Z^d\times \Z^d}$, 
if all these constants are finite. 
\end{defn}
For this class of interactions it is not hard to see from \eqref{stdphi} that 

\begin{equation}\begin{split}\label{lippi1}
{\rm std}_{i,j}(\P)\leq L_{ij}\sup_{\eta_i\in S'}\inf_{a_i\in \S^n}\Bigg(\int \a_{\eta_i}(d\s_i)d(\s_i,a_i)^2\Bigg)
^{\frac{1}{2}}.
\end{split}
\end{equation}
This follows from replacing the $b$ in \eqref{stdphi} by $H_i^{\z}(a_i)-H_i^{\bar\z}(a_i)$. 
Let us now see some concrete examples.


\subsubsection{Short-time Gibbsianness of $n-$vector lattice models under diffusive time-evolution}\label{stimegibbs} 
To a Gibbs measure $\mu$ for a lattice interaction $\P$ we apply sitewise independent diffusive dynamics given by 
\begin{equation}\label{kernel}
 K(d\s_i,d\eta_i)=K_t(d\s_i,d\eta_i)=k_t(\s_i,\eta_i)\a_0(d\s_i)\a_0(d\eta_i).
  \end{equation}
In the above $\a_0$ is the equidistribution on $\S^n$ and $k_t$ is the heat kernel on the sphere, i.e.
\begin{eqnarray}
\Big(e^{\D t}\varphi\Big)(\eta_i)=\int \a_o(d\s_i)k_t(\s_i,\eta_i)\varphi(\s_i),
\end{eqnarray} 
where $\D$ is the Laplace-Beltrami operator on the sphere and $\varphi$ is any test function. 
$k_t$ is also called the {\em Gauss-Weierstrass kernel}. 
For more background on the heat-kernel on Riemannian manifolds, see the introduction 
of \cite{ACDH04}. 
Let $\bar C(t)$ be the matrix with entries 
\begin{equation}
\bar C_{i,j}(t)= \frac{L_{ij}}{\sqrt{2}}\bigl(1-e^{-nt}\bigr)^{\frac{1}{2}}\exp\Bigr(\sum_{A\supset
\{i,j\}}\frac{\delta(\Phi_A)}{2}\Bigl).
\end{equation}
With the above notation we have the following generalization of Theorem 2.7 of \cite{KULOP}.

\begin{thm}\label{genthm}
Suppose $d$ is the Euclidean metric on $\S^n$ and $\P$ is an interaction for which there are finite constants 
$(L_i=L_i(\P))_{i\in \Z^d}$ such that 
\begin{equation}\label{lip}
\sup_{\o\in\O}\big|H^\o_i(\s_i)-H^\o_i(a_i)\big|\leq L_i d(\s_i,a_i).
\end{equation}
Assume further that $\sup_{i\in\Z^d}\sum_{j\in \Z^d} \bar C_{ij}(t)<1$. 
Then the transformed measure $\mu_t$ obtained from a Gibbs measure $\mu$ for $\P$ and $K_t$ 
is Gibbsian and the specification for $\mu_t$ has goodness $(\bar Q,d)$ with
\begin{equation}\begin{split}\label{barq}
& \bar Q_{i j}(t):= \frac{1}{2}\min\Bigl\{ 
\sqrt{\frac{\pi}{t}}
Q_{i j}(t), e^{4 L_i}-1\Bigr \}.
\end{split}
\end{equation}
Here $Q(t)$ is defined in the same way as in \eqref{dino} but has $\bar{C}$ replaced by $\bar{C}(t)$.
\end{thm}
The inequality \eqref{lip} implies \eqref{lippi}. But using \eqref{lippi} gives a better bound on the 
Dobrushin interdependence matrix. In \eqref{lip} we keep all the interactions a given site $i$ has with 
the rest of its environment, but in \eqref{lippi} only the interaction between $i$ and a reference site $j$ 
is kept. Note also that the entries of $\bar{C}(t)$ will be small if either the initial interaction is weak or 
$t$ is small enough. This is a generalization of the corresponding Ising and planar rotor results 
found in \cite{WIO,ACD1} to more general interactions on any $n-$dimensional sphere, subjected to 
infinite-temperature diffusive dynamics.\\\\
The above theorem was proved in \cite{KULOP} for some special pair interaction. The proof 
there followed from three steps, namely: 
1) an application of Theorem \ref{mainthm-lat} to obtain continuity estimates in terms 
of the posterior metric $d'$, 
2) a comparison result between $d'$ and $d$ (the Euclidean metric), see e.g Proposition 2.8 of 
\cite{KULOP} 
and 
3) a telescoping argument over sites in the conditioning. The first two steps hold for any general interaction 
on the sphere. \\

In the third step one uses the continuity property \eqref{lip} to proceed. In particular one replaces the constant 
$c_j$ in inequality (100) of \cite{KULOP} by $L_j$.
\begin{flushright}
$\Cox$
\end{flushright}

In the next subsection we consider another class of transformations which 
was studied in \cite{KULOP}.

\subsubsection{Conservation of Gibbsianness for $n-$vector lattice models under discretisations
(fine local approximations)} \label{fine}
Consider a Gibbs measure $\mu$ for an interaction $\P$ satisfying \eqref{lippi}. Furthermore, partition $\S^n$ into countably many 
pairwise disjoint subsets with non-zero $\a$ measure, indexed by
elements in a countable set $S'$. Thus we have disjoint subsets $S_i$, such that $\S^n=\cup_{i\in S'}S_i$ 
and $\a(S_i)>0$ for all $i\in S'$. 
For each such decomposition of $\S^n$ we define $$\rho=\sup_{i\in S'}{\rm diam}(S_i),$$
where ${\rm diam}(S_i)$ is the diameter of the set $S_i$. We refer to $\rho$ as the fineness of the decomposition.
In this set-up the conditional a priori measure is given by $\a_{\eta_i}(d\s_i)=\frac{\a_{|_{S_{\eta_i}}}}{\a(S_{\eta_i})}$. 
The above decompositions of $\S^n$ define natural maps from the space of probability measures on $\O$ to probability
 measures on $(S')^{\Z^d}$. The question now is: "Which of these maps will lead to a Gibbsian image measure $\mu'$ upon their 
 application to the Gibbs measure $\mu$?" This question is partially answered  in Theorem 2.9 of \cite{KULOP}, which
 we state below. 
 
 \begin{thm}\label{fineapp}
 Suppose $\P$ is as above and 
 
 \begin{eqnarray}\label{rho}
\frac{\rho}{2}\sup_{i\in \Z^d}\sum_{j\in i^c }\exp\Bigr(
\frac{1}{2}\sum_{A\supset\{i,j\}}\delta(\Phi_A)\Bigl)L_{ij} < 1.
\end{eqnarray}
 Then for any Gibbs measure $\mu$ of $\P$ the transformed measure $\mu'$, associated with the decomposition with fineness $\rho$,
 is a Gibbs measure for a Gibbsian specification $\g'$ of goodness $(Q,d_0)$. Here $d_0$ is the discrete 
metric on $S'$ and $Q$ is given in \eqref{dino} 
 with $\bar C$ given by
 
 \begin{eqnarray*}\label{rho1}
\bar C_{ij}=\frac{\rho}{2}\exp\Bigr(
\frac{1}{2}\sum_{A\supset\{i,j\}}\delta(\Phi_A)\Bigl)L_{ij}. 
\end{eqnarray*}
 \end{thm}
Observe from the above theorem that the quantity in \eqref{rho} can be small if either the initial interaction is weak or the fineness 
 $\rho$ of the decomposition is small enough. 
Thus for any strength of the initial interaction, the transformed measures will
remain  Gibbsian if our  decomposition is fine enough. We note that if we make 
a decomposition of the circle into equal intervals, the resulting models 
resembles a clock model.     
On could in fact also apply the theorem starting with discrte spins, such as 
a large-$q$ clock model, but the advantage of considering continuous spin 
spaces is that the theorem can always be applied (in other words, there 
is {\em always} a fine enough decomposition). 

Note also that such a discretisation map has strong 
similarities with fuzzification maps such as have been 
studied for Potts models, see e.g. \cite{MaeVel95,HAGG03}, in which one also 
decomposes the single-site spin space, into a smaller number of fuzzy 
spin values.



\subsection{Large-time results: Conservation, Loss and Recovery of Gibbsianness}

This section deals with what is known about conservation, loss and recovery of the Gibbs property in time-evolved Gibbsian measures of vector models on the lattice $\mathbb{Z}^d$.
The conservation part will focus on large-time results, as the short-time results have been described in the previous section. We will concentrate here on the work done in \cite{WIO, WIO1}.
Moreover we will present some new arguments which extend the results in \cite{WIO1}.

\medskip

In the previous section we defined Gibbs measures, see equation \eqref{Gibbs},
and furthermore we gave an equivalent description which we stated in equation \eqref{quasiloc}. Let us now focus
on the latter expression. In words it says if a measure $\mu$ is Gibbsian, every configuration
$\eta$ is \textbf{good}, in the sense that for every $\eta$ the measure is \\ 
continuous w.r.t. a
change in the conditioning. We referred to this property as the \textbf{quasilocality}
property. But what does it mean for a measure $\mu$ to be \textit{non}-Gibbsian?
Loss of Gibbsianness means essentially the failure of this quasilocality
property. It is enough to find at least one point of essential discontinuity
$\eta^{spec}$ w.r.t. the conditioning, namely a point satisfying
\begin{equation*}
\sup_{\omega, \zeta} |\mu_{\Gamma}(\sigma_0 | \eta^{spec}_{\Lambda \setminus \lbrace 0 \rbrace}\omega_{\Gamma \setminus \Lambda}) - \mu_{\Gamma}(\sigma_0 | \eta^{spec}_{\Lambda \setminus \lbrace 0 \rbrace}\zeta_{\Gamma \setminus \Lambda}) | > \e
\end{equation*}
for $\Gamma \supset \Lambda$, uniformly in $\Lambda \subset \Z^d$, to prove that a measure is
non-Gibbsian.
Physically the failure of quasilocality means the following: \\
The spin at the origin $\sigma_0$ is influenced by far away configurations $\omega_{\Gamma \setminus \Lambda}$ and $\zeta_{\Gamma \setminus \Lambda}$ even when the spins in between 
are frozen in the configuration $\eta^{spec}_{\Lambda \setminus \lbrace 0 \rbrace}$ . For a measure which is Gibbsian, the spin $\sigma_0$ is shielded off from spins far away when intermediate spins are fixed. So there are no hidden fluctuations transmitting information from infinity to the origin. \\
Typically, in the analysis one considers a double-layer system with the initial spin space in the first layer and the transformed system (or image-spin space) in the second layer.
The question of the Gibbsianness of the measure on the second layer then can be shown
to reduce to the question: Is it possible to end up in this particular configuration coming
from one initial Gibbs measure only?
It turns out that if the original spin system \textbf{conditioned} on a particular image spin configuration $\eta^{spec}$ exhibits a phase transition, this implies for $\mu^{t_0}$ that this measure is not Gibbsian. The configuration $\eta^{spec}$ is often called a \textbf{bad configuration}. We want to stress the difference between a phase transition of the initial system and a phase transition of the conditioned double-layer system. Even if the initial system exhibits a phase transition and the time-evolved measure at time $t$ is a Gibbs measure, it means that conditioned at every possible image spin configuration $\eta$ at time $t$, there is no phase transition for the conditioned system. In other words, for every possible $\eta$ there is only one possible initial measure leading to this image spin configuration. Hence, a phase transition of the initial system does not necessarily imply non-Gibbsianness, nor does non-Gibbsianness imply a phase transition for the initial measure. \\

In the case of time-evolved XY-spins on the lattice $\Z^d$, in \cite{WIO1} and \cite{WIO}
some results about conservation, loss and recovery of Gibbsianness
were proved. In \cite{WIO} results are proven for conservation and loss of Gibbsianness during
time-evolution. In particular, loss of Gibbsianness could be proven for zero initial external fields.
The paper \cite{WIO1} deals with loss and recovery of Gibbsianness in a situation where there is a positive initial external field. As we already discussed in the previous section,
the Gibbsian property is conserved for short times for all initial Gibbs measures evolving
under diffusive dynamics consisting of Brownian motions on the circle, either with or without
gradient Hamiltonian drifts, at all temperatures (for all values of $T_1$ and all values of the initial temperature $T_2$). Moreover, conservation for all times holds if the system starts with a
high-or infinite-temperature Gibbs measure and evolves under high- or infinite-temperature
dynamics ($T_1 >> 1$). \\
Let us make the statement on the loss of Gibbsianness result from
\cite{WIO} more precise. Systems in $\mathbb{Z}^2$ are considered which start in a low-temperature initial measure with nearest neighbour interaction and zero external field,
\begin{equation*}
\overset{\sim} \varphi_{\Lambda}(\sigma) = - J \underset{i,j \in \Lambda: i \sim j} \sum \cos(\sigma_i-\sigma_j)
\end{equation*}
and evolve under independent Brownian motion dynamics on the circle. Then there is a time interval, depending on the initial (inverse) temperature, such that the time-evolved measure is not Gibbsian. The idea of the proof is the following. Consider the double-layer measure and condition it on the alternating North-South configuration. The ground states of the conditioned system then are two symmetric configurations of spins pointing either to the East with a small correction $\pm \e_t$ or to the West with a small correction. Let us give a schematic picture.

\begin{minipage}[hbt]{5cm}
\centering
\includegraphics[width= 4 cm, height= 4 cm, angle= 270]{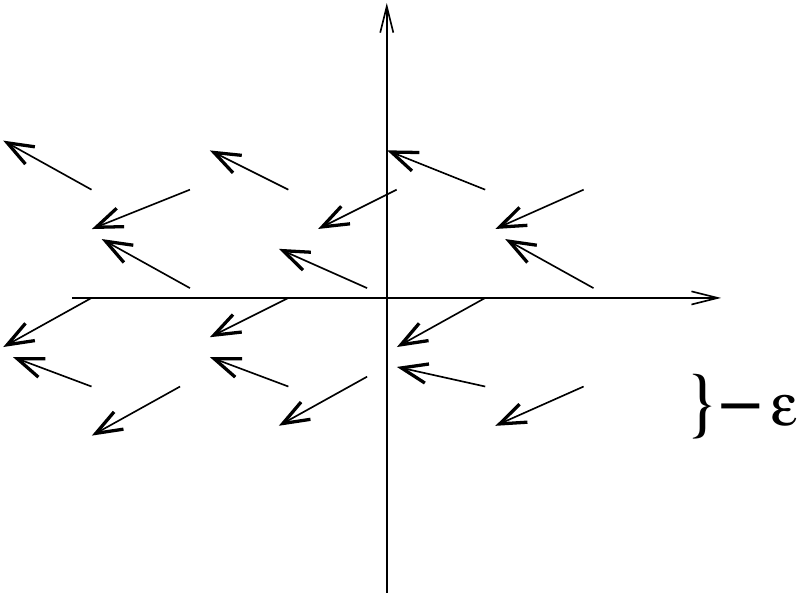}
West-pointing ground state
\end{minipage}
\hfill
\begin{minipage}[hbt]{5cm}
\centering
\includegraphics[width= 4 cm, height= 4 cm, angle= 270]{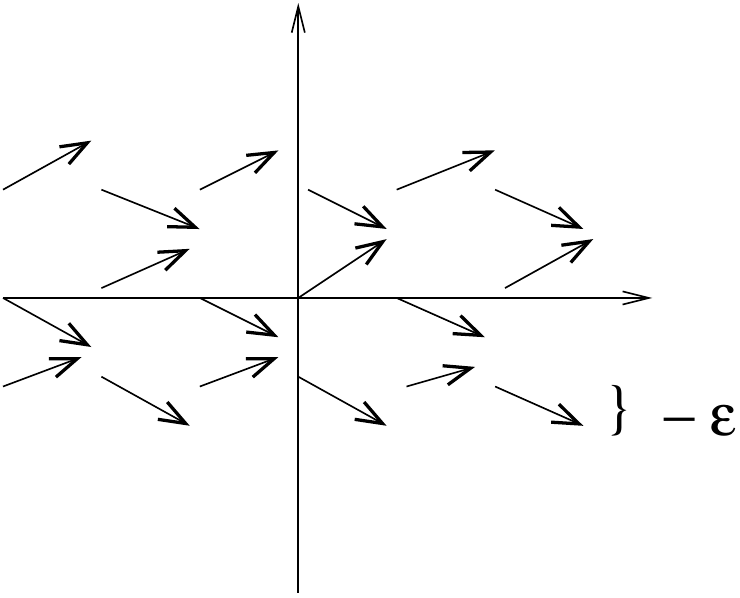}
East-pointing ground state
\end{minipage}

\medskip

The potential of the conditioned model is of C-type (nearest neighbour, invariant under reflection in vertical and horizontal direction and invariant under translation), so that the corresponding measure is reflection positive. Using a percolation argument for low-energy clusters (clusters of vertices connected by low-energy bonds) from \cite{Geo81}, in two dimensions one proves that at low
temperatures there exist two distinct extremal Gibbs measures for the
thus-conditioned system.
This implies that the conditioned double-layer system undergoes a phase transition via discrete symmetry breaking and therefore the time-evolved measure is not Gibbsian.
This phase transition is called of ``spin-flop" type. Let us remark on a special feature of this result. In the Ising spin case see \cite{ACD1}, where one also finds at zero external field loss of Gibbsianness, the initial system itself is already not unique. The XY spin model, however, does not have a first order phase transition in two dimensions. So even though starting from a
unique Gibbs measure, there is a time interval where Gibbsianness is lost.\\
We also mention that the result can be extended to $\Z^3$ and arbitrary large times. In that case the initial measure is not unique, and there is long-range order for any strength of the alternating magnetic field, including zero.

Unfortunately, the techniques which are used rely on the reflection positivity property of the measure, and therefore cannot be applied to a system which evolves with high-temperature dynamics, since then the conditioned measure is not
reflection positive any more. Also for higher-component spins the proof breaks down. \\

For discrete spins, the authors in \cite{ACD1} prove that loss of Gibbsianness appears also for high-temperature dynamics, for rotor spins we believe the same is true but this has not yet been
proven. By some Pirogov-Sinai type arguments one might hope to extend the above result to high-temperature dynamics. But this seems a technically hard question. \\
In the presence of an initial external field $h$ loss and also recovery results were obtained in \cite{WIO1}. Similar to the situation for discrete spins in \cite{ACD1}, one finds that also in the presence of a small initial external field there can be a time interval $(t_0,t_1)$ where Gibbsianness is lost in $d \geq 3$ dimensions. Moreover there exists a time $t_2$ such that for all $t \geq t_2$, the time-evolved measure is again a Gibbs measure. This re-entrance result was obtained for strong initial external fields in $d \geq 2$ lattice dimensions. \\
Intuitively, for an intermediate time interval, the strength of the initial field is compensated by the induced field coming from the time-evolution. This compensation makes the system behave like in a modified zero field situation. The system looks like a zero field system plus some rest terms which have a discrete symmetry instead of a continuous one. For low enough initial temperatures there is a time interval where this symmetry is broken for the conditioned double-layer
system, so therefore Gibbsianness is lost. After some time the influence of the time-induced fields decreases and the system follows the initial field again which brings it back to the Gibbsian regime. Thus the presence of the initial external field is responsible for the recovery of the Gibbsian property. \\
The proof in \cite{WIO} is similar to the one in \cite{WIO1}. One considers a double-layer system and conditions on spins pointing all southwards. Then the two ground states of the conditioned system are again symmetric, pointing either to the East or to the West. We present a schematic picture of the ground states.

\begin{minipage}[hbt]{5cm}
\centering
\includegraphics[width= 4 cm, height= 4 cm]{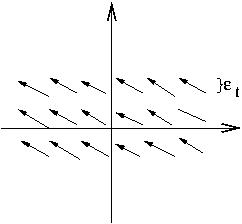}
East-pointing ground state
\end{minipage}
\hfill
\begin{minipage}[hbt]{5cm}
\centering
\includegraphics[width= 4 cm, height= 4 cm]{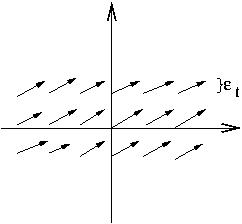}
West-pointing ground state
\end{minipage}

Since the interaction of the conditioned system is also of C-type, i.e. among other properties invariant under reflections, one can use again the percolation of low-energy clusters argument of Georgii, see \cite{Geo81}.\\

The proof for the \textbf{loss} of Gibbsianness works only for $d \geq 3$ and for small initial fields. \\

In the following we propose an argument for a loss of Gibbsianness result which works for a general initial field $h$ already in a two-dimensional lattice.
Moreover we will prove a recovery result valid for all strengths of the initial field at low enough temperatures.

\begin{prop}
Let $h$ be given. For $\beta$ big enough, there exists a time interval \\
$(t_0(\beta, h), t_1(\beta, h))$ such that for $t_0 < t < t_1$, the time-evolved measure $\mu^t$ is not Gibbsian.
\end{prop}
\textbf{Proof:} \\
Let us consider the double-layer joint measure at time 0 and time $t$
simultaneously.
The dynamical Hamiltonian $\textbf{H}^t_{\beta}(\sigma,\eta)$ (the inverse temperature
is in this case absorbed into the definition of the Hamiltonian) is formally
given by
\begin{equation*}
-\textbf{H}^t_{\beta}(\sigma,\eta) = - \beta \overset{\sim}H(\sigma) + \sum_{i \in \Z^2} \log(p_t^{\odot}(\sigma_i,\eta_i)),
\end{equation*}
where $\sigma, \eta \in [0,2\pi)^{\Z^2}$, $p_t^{\odot}(\sigma_i,\eta_i)$ is the transition
kernel on the circle and the initial Hamiltonian $\overset{\sim}H(\sigma)$ is formally given by
\begin{equation*}
-\overset{\sim}H(\sigma) = J \sum_{i \sim k} \cos(\sigma_i-\sigma_k) + h\sum_i \cos(\sigma_i)
\end{equation*}
while $p_t^{\odot}$ equals
\begin{equation*}
p_t^{\odot}(\sigma_i,\eta_i) = \frac{1}{\sqrt{2\pi t}} \sum_{n \in \mathbb{Z}} e^{-\frac{(\sigma_i-\eta_i + 2\pi n)^2}{2t}}.
\end{equation*}

We condition the system at time $t$ to point down alternatingly with a small correction $\varepsilon$ (which we will specify later) to the East,
resp. with a small correction to the West,  i.e.
\begin{equation*}\label{1}\begin{split}
&\eta^{spec}_{i , \varepsilon}= \begin{cases}
  \pi + \varepsilon  & \mbox{if } |i| = i_1 + i_2 \text { is even} \\
  \pi - \varepsilon & \mbox{else}
\end{cases}
\end{split}
\end{equation*}

Let us look at the corresponding dynamical Hamiltonian.
It can be written as a sum over all nearest neighbor
pairs of the pair interaction-potential
\begin{equation*}
\begin{split}
&\textbf{H}^t_{\beta, \lbrace i, i+1 \rbrace}(\sigma,\eta^{spec}_{\varepsilon}) = \cr
& \Phi_{\beta}^{t,\varepsilon}(\sigma_{i}, \sigma_{i+1}) = -\beta J \cos(\sigma_i-\sigma_{i+1}) - \beta \frac{h}{4}\biggl(\cos(\sigma_i)+\cos(\sigma_{i+1}) \biggr)  \cr
& -\frac{1}{4}\log(p_t^{\odot}(\sigma_i,\pi +\e  ))- \frac{1}{4}\log(p_t^{\odot}(\sigma_{i+1},\pi - \e  ))
\end{split}
\end{equation*}
where $h, \b, t, \e$ are parameters (at $J$ fixed), and $\sigma_i, \sigma_{i+1}$ denote the values of neighboring spins.
The single-site terms coming from the dynamics play the role of dynamical alternating external fields. Let us rewrite the terms as
\begin{equation*}
\begin{split}
&\log(p_t^{\odot}(\sigma_i,\pi +\e  )) = \cr
& -\log(\sqrt{2\pi t}) - \frac{(\sigma_i -( \pi + \e) )^2}{2t} + \log \biggl( 1 + \sum_{n \in \Z \setminus 0}e^{-\frac{(\sigma_i - (\pi + \e) + 2n\pi)^2}{2t} + \frac{(\sigma_i -( \pi + \e) )^2}{2t}} \biggr )
\end{split}
\end{equation*}
and similarly for the second one. We choose $t$ of order $\mathcal{O}(1/\beta h)$, or more precisely in such a way that
\begin{equation*}
\begin{split}
& \beta h \biggl(\cos(\sigma_i)+\cos(\sigma_{i+1}) \biggr) = \cr
& 2\log(\sqrt{2\pi t}) + \frac{(\sigma_i -( \pi + \e) )^2}{2t} + \frac{(\sigma_{i+1} -( \pi - \e) )^2}{2t} + o_t(\sigma_i^4, \sigma^4_{i+1})
\end{split}
\end{equation*}
where $o_t(\sigma_i^4, \sigma^4_{i+1})$ is an error term going to 0 for $t$ small.
Call
\begin{equation*}
R_t(\sigma_i, \pi + \e):=\log \biggl( 1 + \sum_{n \in Z\setminus 0}e^{-\frac{(\sigma_i - (\pi + \e) + 2n\pi)^2}{2t} + \frac{(\sigma_i -( \pi + \e) )^2}{2t}} \biggr )
\end{equation*}
and similarly for $R_t(\sigma_{i+1}, \pi - \e)$. Observe that the pair interaction potential is equal to
\begin{equation*}
\Phi_{\beta}^{t,\varepsilon}(\sigma_{i}, \sigma_{i+1}) = -\beta J \cos(\sigma_i-\sigma_{i+1})
-\frac{1}{4}R_t(\sigma_i, \pi + \e)- \frac{1}{4}R_t(\sigma_{i+1}, \pi - \e) + o_t(\sigma_i^4, \sigma^4_{i+1}).
\end{equation*}
We end up with a ferromagnetic system with alternating dynamical external fields of strength $\mathcal{O}(\e/t)$ coming from the terms $R_t(\sigma_i, \pi + \e)$ which effectively point perpendicular to the original fields. Let us assume $\e/t << \beta J$. Then the strength of the fields is in fact of order $\mathcal{O}(\e h/J)$ and and the direction alternates, pointing almost to the East or almost to the West while the strength is relatively weak compared to the nearest-neighbour interaction. We will be able to show that the spin flop mechanism causes a phase transition to occur. \\
In order to understand the phases of such a Hamiltonian we will look at first at its ground states. We want that $\Phi_{\beta}^{t,\varepsilon}(\sigma_{i}, \sigma_{i+1})$ is minimal at $(\delta_t,-\delta_t)$ and $(\pi + \delta^{\prime}_t,\pi -\delta^{\prime}_t)$, so the ground states point almost to the North, namely in North-East and North-West direction, or almost to the South. In general one of them is a local minimum and one is global. One determines $\delta_t$ and $\delta_t^{\prime}$ in such a way that the configurations $(\delta_t,-\delta_t)$ and $(\pi + \delta^{\prime}_t,\pi -\delta^{\prime}_t)$ are the only minima. To make them both global, we specify $\e=\e(h,t)$ such that the following equation is true
\begin{equation*}
\begin{split}
0 & =  \Phi_{\beta}^{t,\varepsilon}(\delta_t, -\delta_t) -  \Phi_{\beta}^{t,\varepsilon}(\pi + \delta^{\prime}_t, \pi - \delta^{\prime}_t) \cr
& = -\frac{1}{4}\biggl (R_t(\delta_t, \pi + \e) - R_t(\delta^{\prime}_t, \e) + R_t(-\delta_t, \pi - \e) - R_t(-\delta^{\prime}_t, \e) \biggr ).
\end{split}
\end{equation*}
In contrast to the zero-field situation,
the spin flip between $\sigma_i$ and $\pi -\sigma_i$ is not a symmetry of the Hamiltonian any more. Indeed, for the particular choice of the time $t$ and $\e$ two ground states occur which are not related by any symmetry. As we described above, the conditioning more or less cancels in the direction of the original field, and one ends up with a model having alternating single-site terms (external
fields), which are pointing almost perpendicular to the original fields.
The coexistence of the ground states can, for temperature T low enough, be
extended to coexistence of two Gibbs measures, now not related by any spin-flip
symmetry, for a slightly different choice of $\e=\e(h,t)$. As the choice of
the ``bad'' conditioning configuration which contains the
$\e(h,t)$ can be continuously varied,
we can deduce the existence of a time-interval of non-Gibbsianness.\\
We remark that, unlike the two previous cases where one could use the
reflection positivity property of the measure (as well as the spin-flip
symmetry) in this case unfortunately we cannot. The interaction is not
invariant under lattice reflections, so the measure is not reflection
positive. We have to use other techniques. We will use a general contour
argument from \cite{Mal83}. Let us recall the statement.
\begin{thm}[Theorem 6 from \cite{Mal83}]
Let $S=[0,1] \subset \R$ and let $\Psi(s_1,s_2,u_1,...,u_{N-1})$ be an (N-1)-parameter family of potentials defined for $u=(u_1,...,u_{N-1}) \in \R^{N-1}$ varying in a neighbourhood of 0 in $\R^{N-1}$. Assuming the family $\Psi(s_1,s_2,u)$ satisfies the following conditions
\begin{enumerate}
\item the function $\Psi(s_1,s_2,u)$ is smooth in all its variables
\item for $u=0$ the function $\Psi(s_1,s_2,0)$ has N absolute minima at points situated on the diagonal of the square $S \times S$, i.e.
\begin{eqnarray*}
\Psi(m_i,m_i,0)=0 & & \text{ for all } i \\
\Psi(s_1,s_2,0) > 0 & & \text{ for all } (s_1,s_2) \neq (m_i,m_i)
\end{eqnarray*}
\item at the minima $(m_i,m_i)$ the second differential of the function $\Psi(s_1,s_2,0)$ is strictly positive and moreover
\begin{equation*}
\biggl| \frac{d^2 \Psi}{d s_1 d s_2} \biggl|_{s_1=s_2=m_i} \biggr| < \eta \frac{d^2 \Psi}{d s_1^2} \biggl|_{s_1=s_2=m_i}
\end{equation*}
where $\eta$ is a sufficiently small constant,
\item at points $(m_i,m_i)$, the differentials of $\Psi(s_1,s_2,u)$ at $u=0$ are nonzero,
\end{enumerate}
THEN there exists a point $u_0=u_0(\beta)$ such that for the system described by the potential $\Psi(s_1,s_2,u_0)$ there exist at least N distinct limit Gibbs distributions.

\end{thm}
We want to apply the above Theorem. For the assumptions to be satisfied we
have to transform and shift our potential $\Phi_{\beta}$, to apply the statement about translation invariant potentials to a statement which also applies to periodic ones.
We will define our new potential $\Psi$ on $S^2 \times S^2$ instead of $S \times S$ as required in the assumptions. This does not affect the proof in any essential way. Our spin space $\mathbb{S}^1$ is isomorphic to $[0,1]$ by the isomorphy $\sigma \mapsto \sigma/2\pi$, where 0 and 1 are considered to be the same points. We abbreviate $\sigma^{\prime} := \sigma/2\pi$. Let $u$ be a smooth function around a small neighbourhood of 0 in $\R$ and let $m:= \inf_{\sigma,\zeta} \lbrace \Phi^{t,\e}_{\beta}(\sigma,\zeta) \rbrace$. We define the new potential $\Psi^{t,\e}_{\beta}(\sigma^{\prime}_1, \zeta^{\prime}_1, \sigma^{\prime}_2, \zeta^{\prime}_2,u)$ as being a sufficiently smooth function of all its variables. Furthermore let the differentials of $\Psi^{t,\e}_{\beta}(m_i,m_i,u)$ at $u=0$ be nonzero. For $u=0$ the potential is given by
\begin{equation}
\Psi^{t,\e}_{\beta}(\sigma^{\prime}_1, \zeta^{\prime}_1, \sigma^{\prime}_2, \zeta^{\prime}_2,0):= \Phi^{t,\e}_{\beta}(\sigma^{\prime}_1,\zeta^{\prime}_1) + \Phi^{t,\e}_{\beta}(\sigma^{\prime}_2,\zeta^{\prime}_2) - 2m ,
\end{equation}
note that it is isomorphic to $\Phi^{t,\e}_{\beta}$.
Then obviously $\Psi^{t,\e}$ inherits the two minima from $\Phi^{t,\e}_{\beta}$ which we call $m_1$ and $m_2$.
The second assumption is satisfied by the definition of $\Psi^{t,\e}$.
Let us further examine the determinant of the Hessian matrix to check the third condition. We call $Hess(\Psi^{t,\e}_{\beta}(\sigma^{\prime}_1, \zeta^{\prime}_1, \sigma^{\prime}_2, \zeta^{\prime}_2,0))$ the Hessian matrix for the function $\Psi^{t,\e}_{\beta}$. Then one observes that for the determinant of the Hessian
\begin{eqnarray*}
\det( Hess(\Psi^{t,\e}_{\beta}(\sigma^{\prime}_1, \zeta^{\prime}_1, \sigma^{\prime}_2, \zeta^{\prime}_2,0))) & = & \det(Hess(\Phi^{t,\e}_{\beta}(\sigma^{\prime}_1, \zeta^{\prime}_1))) \det(Hess(\Phi^{t,\e}_{\beta}(\sigma^{\prime}_2, \zeta^{\prime}_2))) \\
& = & \biggl ( \det (Hess(\Phi^{t,\e}_{\beta}(\sigma^{\prime}_1, \zeta^{\prime}_1))) \biggr )^2
\end{eqnarray*}
which is strictly positive at the minimal points $(m_1,m_1)$ and $(m_2,m_2)$ for the parameters $t$ and $\e$ chosen above and $\beta$ big enough.
Then using the theorem we deduce that for a sufficiently large $\beta$ there exists a $u_0$ such that for the system described by $\Psi^{t,\e}_{\beta}(\sigma^{\prime}_1, \zeta^{\prime}_1, \sigma^{\prime}_2, \zeta^{\prime}_2,u_0)$ there exist at least 2 distinct Gibbs measures. Since $\Psi^{t,\e}_{\beta}(\sigma^{\prime}_1, \zeta^{\prime}_1, \sigma^{\prime}_2, \zeta^{\prime}_2,u_0)$ and $\Psi^{t,\e}_{\beta}(\sigma^{\prime}_1, \zeta^{\prime}_1, \sigma^{\prime}_2, \zeta^{\prime}_2,0)$ are physically equivalent this follows also for our potential $\Phi^{t,\e}_{\beta}$.

\begin{flushright}
$\square$
\end{flushright}

Let us now present  a recovery result which will be valid
for all strengths of the initial field at sufficiently low temperatures.

\begin{prop}
Let $h$ be given, then for $t$ large enough and $\beta$ large enough, for example of order $\mathcal{O}(e^{t^2})$ , there is a unique time-evolved measure.
\end{prop}
\textbf{Proof:}\\
We want to prove that for large enough times the time-evolved measure is unique. To prove this we want to use Theorem 7 from \cite{Mal83} which is a Pirogov-Sinai type argument for continuous models with one ground state. Let us cite their Theorem 7.

\bigskip

\begin{thm}[Theorem 7 from \cite{Mal83}]
Let $S=[-1,+1] \subset \mathbb{R}$ and let us consider the lattice $\mathbb{Z}^d$. Suppose that the function $\Psi(s_1,s_2)$ is smooth in a neighbourhood of $(0,0)$ and on $S \times S$ achieves an absolute minimum at $(0,0)$. Let us also suppose that $\Psi(0,0)=0$. Moreover let the expansion of
$\Psi(s_1,s_2)$ in a neighbourhood of $(0,0)$ have the form
\begin{equation}
\Psi(s_1,s_2) = s_1^2 + 2\eta s_1s_2 + s_2^2 + \mathcal{O}(s_1^3 + s_2^3)
\end{equation}
where $\eta$ is a small (positive OR negative constant).

\medskip
THEN there exists a temperature $\beta_0 = \beta_0(\Psi,d)$ such that for $\beta > \beta_0$ there exists a unique limit Gibbs distribution which depends analytically on $\beta$.
\end{thm}

\bigskip

$S$ is the state space of the spins and $\Psi$ is the potential on the product space $S \times S$. So all we have to do is again rewrite our potential and prove the assumptions given in the Theorem.
Our original potential without approximation is given by
\begin{equation*}\begin{split}
\Phi_{\beta}(\sigma_{i},\zeta_{i+1}) &= - \beta J \cos(\sigma_{i}-\zeta_{i+1}) - \frac{\beta h}{4}\biggl( \cos(\sigma_{i})+\cos(\zeta_{i+1}) \biggr)\cr & - \frac{1}{4} \biggl (\log(p_t^{\odot} (\sigma_{i},\eta_i))- \log(p_t^{\odot} (\zeta_{i+1} ,\eta_{i+1}) ) \biggr ).
\end{split}
\end{equation*}
It is defined including the inverse temperature $\beta$, which does not pose a problem. For $t$ large enough the unique minimum of $\Phi_{\beta}$ is equal to $(0,0)$. Let us rescale the potential $\Phi_{\beta}(\sigma_{i},\zeta_{i+1})$ by $\sigma^{\prime}: \sigma \mapsto \sigma/2\pi$ and consider the isomorphy $[0,2\pi]/2\pi \simeq [-1,1]$ where $-1$ and $1$ are considered to be the same points. Moreover we subtract a constant from the potential to assure that $\Phi_{\beta}(0,0)=0$, i.e.
\begin{equation*}\begin{split}
& \Phi^{\prime}_{\beta}(\sigma_{i},\zeta_{i+1}) = - \beta J \cos (\sigma^{\prime}_{i}-\zeta^{\prime}_{i+1}) - \frac{\beta h}{4}\biggl(\cos(\sigma^{\prime}_{i})+\cos(\zeta^{\prime}_{i+1}) \biggr) - \frac{1}{4} \log(p_t^{\odot} (\sigma^{\prime}_{i} ,\eta^{\prime}_i ))\cr &- \frac{1}{4}\log(p_t^{\odot} (\zeta^{\prime}_{i+1} ,\eta^{\prime}_{i+1} )) +
  \beta J + \beta h/2 + \frac{1}{4}\log(p_t^{\odot}(0,\eta^{\prime}_i)) + \frac{1}{4}\log(p_t^{\odot}(0,\eta^{\prime}_{i+1}))
  \end{split}
\end{equation*}
We call $f^t(\sigma^{\prime}_{i},\eta^{\prime}_i):= \frac{1}{4}\log(p_t^{\odot} (0,\eta^{\eta}_i )) - \frac{1}{4}\log(p_t^{\odot}(\sigma^{\prime}_{i} ,\eta^{\prime}_i))$ and write the above potential as
\begin{equation*}\begin{split}
\Phi_{\beta}^{\prime}(\sigma^{\prime}_{i},\zeta^{\prime}_{i+1})& = - \beta J \biggl [ \cos ( \sigma^{\prime}_{i}-\zeta^{\prime}_{i+1}) - 1 \biggr ] - \frac{\beta h}{4}\biggl [ \cos(\sigma^{\prime}_{i} ) -1 +\cos(\zeta^{\prime}_{i+1}) -1 \biggr ]\cr & + f^t(\sigma^{\prime}_{i},\zeta^{\prime}_i) + f^t(\zeta^{\prime}_{i+1},\eta^{\prime}_{i+1}).
\end{split}
\end{equation*}
Around the absolute minimum $(0,0)$, we have the following expansion of $\Phi_{\beta}^{\prime}(\sigma^{\prime}_{i},\zeta^{\prime}_{i+1}) $, using the abbreviation $c(J,h):= \frac{4J + h}{4 (2\pi)^2}$ :
\begin{equation*}\begin{split}
\Phi_{\beta}^{\prime}(\sigma^{\prime}_{i},\zeta^{\prime}_{i+1})& = \beta c(J,h) (\sigma^{\prime}_{i})^2 + \beta \biggl( \frac{-2J}{(2\pi)^2} \biggr )\sigma^{\prime}_i\zeta^{\prime}_{i+1} + \beta c(J,h) (\zeta^{\prime}_{i+1})^2\cr & + \mathcal{O}\biggl((\sigma^{\prime}_{i})^4 + (\zeta^{\prime}_{i+1})^4 \biggr) + o_{i, i+1}(e^{-t}).
\end{split}
\end{equation*}
We  clearly  see that the expansion gives us for $t$ large enough, at least bigger than $\log(c(J,h))$, the desired quadratic form required for Theorem 7.
\begin{flushright}
$\square$
\end{flushright}

\section{Gibbsianness of $n-$vector mean-field models and their transforms}
Mean-field models are spin systems whose distribution is permutation-invariant. 
In \cite{KUL1,KUL2,HaggKu04} the Gibbs properties 
of various mean-field models (with finitely many spin values) were investigated when 
subjected to various transformations. 
In the recent study in \cite{KULOP08}, extensions  
to more general mean-field models with compact Polish spaces as their single-site spin space are discussed. 
We describe these results in this section, restricting to the case where the 
spins take values on a sphere.
Let us start by recalling the general notion of mean-field models and what it takes for them to be Gibbsian.

 
\subsection{General Mean-field Models and Mean-Field Gibbsianness} 
We now present the definition of general mean-field models and mean-field Gibbsianness for such models 
 for $n-$vector spins \cite{KULOP08}. 
 We write $V_N=\{1,2,\dots,N\}$ for the volume at size $N$. 
 
\begin{defn}\label{mean-field-model}
For each $N\in\N$, let $\mu_N$ be a probability measure on the space $(\S^n)^N$. 
\begin{enumerate}
\item We refer to the sequence of the probability measures $(\mu_N)_{N\in\N}$ as a \textbf{mean-field model} if 
the $\mu_N$'s are permutation invariant.
\item A mean-field model $(\mu_N)_{N\in\N}$ is said to be Gibbsian if the following holds:
\begin{enumerate}
\item The limit of conditional probabilities 
 \begin{eqnarray}
\g_1(dx_1|\l):=\lim_{N\uparrow\infty}\mu_{N}\left(d x_1\big| x^{N}_
{V_{N}\ba 1}\right),
\end{eqnarray} 
exists for any sequence $x^{N}_{V_{N}\ba 1}=(x^{N}_{i})_{i \in V_N \ba 1}$ of conditionings  
for which the empirical distribution converges weakly as $N$ tends to infinity,   
$\l=\lim_{N\uparrow\infty}\dfrac{1}{N}\sum_
{i=2}^N\delta_{x^N_i}$.
\item The 
function $\l\mapsto\g_{1}(\cdot |\l)$ is weakly continuous. 
\end{enumerate}
\end{enumerate}
\end{defn}
 
In the above, Gibbsianity of mean-field models is 
defined in terms of the asymptotic behavior of a 
sequence of measures instead of a single limiting measure. 
This is in contrast to the 
lattice case where we just investigated the single infinite-volume measure. 
The results one would get by only looking at the 
infinite-volume limit measures for mean-field systems would provide a lot less, and in some sense 
misleading, information. Indeed, such limit measures are either product measures, and thus trivially Gibbsian, or nontrivial mixtures of product measures and thus highly non-Gibbsian (see for this fact \cite{ELo}).   


The notion of Gibbsianness given in Definition \ref{mean-field-model}
 is equivalent to the one considered in \cite{KUL1,KUL2} for the corresponding Curie-Weiss model (for which of course one has a simpler single-site spin space and measure). This 
 is the case since the 
distribution of a binary random variable is uniquely determined by its mean. Hence for the Curie-Weiss model 
conditioning on the empirical averages gives the same information as conditioning on empirical measures.
For the rest of this section the term  ``Gibbsian'' should be taken in the sense provided 
by Definition \ref{mean-field-model}.

\subsubsection{Mean-Field Interactions}
In the above we have defined general mean-field models. We are now going to prescribe a class of mean-field 
models given via some potential functionals defined on the space of measures on the single-site spin space 
introduced in \cite{KULOP08}. 
In the following we have denoted by $\MM(\S^n)$ and $\MM_+(\S^n)$ the spaces of finite signed measures and finite 
measures on $\S^n$ respectively.

\begin{defn}\label{mfinteraction}
We shall refer to a map $\P: {\MM}_+(\S^n)\rightarrow \R$ as a proper mean-field
 interaction if:
\begin{enumerate}
\item it is weakly continuous,
\item it satisfies the uniform directional differentiability condition, 
meaning that, for each $\nu\in {\MM}_+(\S^n)$ the derivative 
$\P^{(1)}(\nu,\mu)$ at $\nu$ in direction $\mu$ exists and we have 
\begin{equation}\label{differentiability}
\P(\nu+\mu)-\P(\nu)-\P^{(1)}(\nu,\mu)=r(\mu)
\end{equation}
with
 $\lim_{t\rightarrow 0^+}
\frac{r(t\mu)}{t}=0$ uniformly in $\mu\in\MM(\S^n)$ for which $\nu+t\mu\in\MM_+(\S^n)$, 
for $t\in[0,1]$ and
\item $\P^{(1)}(\nu,\mu)$ is a continuous function of $\nu.$ 
\end{enumerate}

\end{defn}
Standard examples are the quadratic interactions for the $q-$state mean-field Potts and the 
Curie-Weiss model (which are defined on the product spaces of finite 
single-site spin spaces, instead of $n$-spheres, with symmetric a priori measure) 
and which are respectively given by
\begin{equation}\label{potts-cw}
\P^P(\nu)=-\frac{1}{2}\sum_{i=1}^q \nu(i)^2\quad \text{and}\quad \P^{CW}(\nu)=-\frac{1}{2}m(\nu)^2,
\end{equation}
where $m(\nu)$ is the mean of the measure $\nu$.

For each mean-field interaction $\P$ and each $N\in\N$ we define the \textbf{finite-volume Hamiltonian} 
$H_N$ (a real-valued function on the product space $(\S^n)^N$) as
\begin{equation}\label{freebd}
H_N(\s_{V_N}):=N\P\big(L_N(\s_{V_N})\big),
\end{equation} 
where $L_{N}\left(
 \s_{V_N}\right)=\frac{1}{N}\sum_{i=1}^{N}\delta_{\sigma_{i}} $
is the empirical measure of the configuration $\s_{V_N}$. Observe from the permutation 
invariance of the empirical measures that $H_N$ is permutation invariant.
With this notation the sequence of probability measures $\mu_{\b,N}$ 
associated with the finite-volume Hamiltonians $H_N$ at inverse temperature $\beta$ given by
\begin{equation}\label{mean-field}
\mu_{\beta,N}(d\s_{V_N}):=\dfrac{e^{-\beta H_N(\s_{V_N}
 )}\a^{\otimes N}
(d\tilde \s_{V_N})}{\int_{(\S^n)^N}e^{-\beta H_N(\bar\s_{V_N}
 )}\a^{\otimes N}
(d\bar\s_{V_N})}
\end{equation} 
is a mean-field model (on $\S^n$ associated with $\P$ and the a priori measure $\a$). 
In the above we have used $\otimes$ to denote the tensor product of measures.
 In the following, unless otherwise stated, the inverse temperature $\beta$ will be absorbed 
 into the interaction $\P$ and we write $\mu_N$ instead of $\mu_{\beta,N}$. 
 We will, with abuse of notation, write $\mu_N$ for the sequence $(\mu_N)_{N\in\N}$. It is show in Proposition 2.4 
 of \cite{KULOP08} that the mean-field models obtained in this way are 
Gibbsian.
Having disposed of the discussion on Gibbsianness for general $n-$vector mean-field models, we now turn 
 our attention to discussing Gibbs properties of transforms of Gibbsian $n-$vector mean-field models. 

 \subsection{Gibbsianness of Transformed $n-$vector   Mean-Field\\ Models}
We now review the notion of Gibbsianness for transformed Gibbsian $n-$vector mean-field models as found in \cite{
KULOP08}. We take $S'$ as the 
single-site spin space for the transformed system, which we also assume to be a compact complete separable metrizable space. 
Further,
we let $\a'$ be some appropriately chosen a priori probability measure on $S'$. Now we take $K(d\s_i,d\eta_i)$ as the joint 
a priori probability measure on $\S^n\times S'$ given by 
\begin{equation}
\begin{split}
K(d\s_i,d\eta_i):=k(\s_i,\eta_i)\a(d\s_i)\a'(d\eta_i)\in\PP(\S^n\times S'),, 
\end{split}
\end{equation}
where  $$k:\S^n\times S'\rightarrow (0,\infty)$$
just as we had before for the corresponding  transformed lattice spin models.
Now the question of interest is the following. 
Starting with an initial sequence of mean-field Gibbs measures $\mu_N$, associated to a fixed general 
mean-field interaction $\P$, will the 
transformed sequence of measures $\mu'_N$ with $(\a')^N$ density
\begin{equation}
\frac{d\mu'_N}{d(\a')^{N}}(d\eta)=\int_{(\S^n)^N}\prod_{i=1,\dots,N}k(\s_i,\eta_i)\mu_N(d\s) 
\end{equation}
be Gibbsian?  In other words, will the transformed single-site kernel a) exist, and b) 
be a continuous function of  the empirical measures of the conditionings?

It is shown in Theorem 3.10 of \cite{KULOP08} that the transformed mean-field model $\mu'_N$ will remain Gibbsian if a certain 
constrained potential function has unique global minimizer, uniformly over the domain of the constraint variable.
The ideology behind this theorem is the same as in the lattice: absence of hidden phase transition 
in the initial system, constrained to be mapped to a fixed configuration in the transformed system implies 
Gibbsianity for the transformed system. In the mean-field situation estimates can be made explicitly.  
 To see something concrete, the authors in \cite{KULOP08} focused attention on mean-field interactions 
$\P$ of the form
\begin{equation}
\P(\nu)=F(\nu[g_1],\ldots,\nu[g_l]),
\end{equation}
where $g_i$ are some fixed bounded non-constant real-valued measurable functions defined on 
$\S^n$, $l\geq1$ and $F:\R^l\rightarrow \R$ is some twice continuously differentiable 
function. In the above we have denoted by $m_i=\nu[g_i]$ the expectation of $g_i$ with respect to $\nu$.
We further set $F_{j}(m)=\frac{\partial }{\partial m_j}F(m)$  and $F_{ju}(m)=\frac{\partial^2}
{\partial m_j\partial m_u}F(m)$. 

Additionally, we assume that $g=(g_1,\cdots,g_l)$ is a Lipschitz function from $\S^n$ to $\R^l$,
with Lipschitz-norm 
\begin{equation}\begin{split}\label{Lipnorm}
 &\Vert g\Vert_{d,2} =\sup_{\s_i\neq \bar\s_i}\frac{\| g(\s_i)-g(\bar \s_i)\|_2}
 {d(\s_i,\bar\s_i)},
 \end{split}
\end{equation}
where $d$ is the metric on $\S^n$. We also denote by $\d (g)$ the sum of the oscillations of the components of 
$g$, i.e.
\begin{equation}\label{deltag}
\d(g)=\sum_{j=1}^l \sup_{\s_i,\bar\s_i\in \S^n}|g_j(\s_i)-g_j(\bar\s_i)|.
\end{equation}
For any $g$ satisfying the above conditions we set
\begin{equation}\label{dbarg}
D_{g}=\overline{\left\{\nu[g]:\nu\in\PP(\S^n)\right\}}.
\end{equation}
Note that $D_g$ is compact subset of $\R^l$ by the boundedness of $g$.
In the sequel we will write $\|\partial^2F\|_{\text{max},\infty}$ for the supremum of the matrix max norm
of the Hessian $\partial^2 F$. i.e.
\begin{equation}\label{maxnorm}
\|\partial^2F\|_{\text{max},\infty}=\sup_{m\in D_g}\|\partial^2 F\big(m\big)\|_{{\rm max}},\quad 
\text{where}\quad \|\partial^2 F\big(m\big)\|_{{\rm max}}=\max_{1\leq i,j\leq l} \big| F_{ij}(m)\big|.
\end{equation}
Furthermore, we also set 
\begin{equation}
\d_{F,g}=\sup_{m\in D_g}\sup_{\s_i,\bar\s_i\in \S^n}\Big|\sum_{j=1}^l F_j(m)\Big(
g_j(\s_i)-g_j(\bar\s_i)\Big)\Big|.
\end{equation}
With the above interaction, it is proved in \cite{KULOP08} that the transformed system associated to any 
$K(d\s_i,\eta_i)=k(\s_i,\eta_i)\a(d\s_i)\a'(d\eta_i)$ 
will remain Gibbsian if a certain quantity is small. Before we make the above result more precise, some more notation 
is in order. We set 
\begin{equation}\begin{split}
\rm{std}_{\a}(k)&:=\sup_{\eta_i\in S'}\inf_{a_i\in \S^n}\Bigl(\int_{\S^n} d^2(\s_i,a_i)k(\s_i,\eta_i)
\a(d\s_i)\Bigr)^{\frac{1}{2}}\quad\text{and}\cr
C(F,g)&:=2\|\partial^2F\|_{\text{max},\infty}\; \d(g)\Vert g\Vert_{d,2}
\exp\left(\frac{\d_{F,g}}{2}\right).
\end{split}
\end{equation}
With these notation we have the following theorem. 
\begin{thm}\label{generalint}
Consider the transformed system $\mu'_N$ associated to the initial mean-field model $\mu_N$ (given by the interaction $\P$ satisfying the above 
conditions) and joint a priori measure $K$ described above. Suppose that
\begin{equation}\begin{split}
&C(F,g)\rm{std}_{\a}(k)<1.
\end{split}
\end{equation}
Then 
\begin{enumerate}
\item the transformed system is Gibbsian and 
\item the single-site kernel $\g_1'$ of the transformed system satisfies the continuity estimate
\begin{equation}\begin{split}
\|\g_1'(\cdot|\nu'_1)-\g_1'(\cdot|\nu'_2)\|&\leq C(F,g)^2\rm{std}_{\a}(k)\rm{std}_{\a}\|\nu'_1-\nu'_2\|,
\end{split}
\end{equation}
where $\rm{std}_\a=\rm{std}_\a(1)$ and $\|\nu'_1-\nu'_2\|$ is the variational distance between $\nu'_1$ and $\nu'_2$. 
\end{enumerate}
\end{thm}
The above theorem is found in \cite{KULOP08} as Theorem 4.3. The smallness of the quantity $C(F,g)\rm{std}_{\a}(k)$
may come from two sources; namely \\
1) the smallness of $C(F,g)$, arising from the weakness of the interaction $\P$ among 
the components of the initial system \\
and \\
2) the smallness of $\rm{std}_{\a}(k)$, coming from the good concentration property 
of the conditional measures $\a_{\eta_1}(d\s_1):=k(\s_1,\eta_1)\a(d\s_1)$, uniformly in $\eta_1\in S'$. \\
Thus we could 
start with a very strong interaction, but if the measures $\a_{\eta_1}(d\s_1)$ are close to delta measures then the 
transformed system will be Gibbsian.
An advantage of this result is that it provides explicit continuity estimates for $\g_1'$ whenever the transformed 
system is Gibbsian, which were lacking in all the results before.
However, it has the drawback that the estimates it provides for the regions in parameter space where the transformed 
system is Gibbsian might not be sharp, as techniques employed in \cite{HaggKu04} and \cite{KUL2} do provide.

We now review two examples discussed in \cite{KULOP08}, 
which are reminiscent of some of the results found in \cite{HaggKu04,KUL2}.

\subsubsection{Short-time Gibbsianness of $n-$vector mean-field models under diffusive time-evolution}\label{example2}
Here we present the result found in \cite{KULOP08} but for general mean-field interactions $\P$ given in terms of 
$F$ and $g$. We study the Gibbs properties of the transformed (time-evolved) system $\mu'_{t,N}$ obtained upon application of
infinite-temperature diffusive dynamics to the initial Gibbsian mean-field model $\mu_N$, associated with $\P$.
In this set-up $S'=\S^n$. The joint single-site a priori measure $K$ is then given as in \eqref{kernel} of Subsection 
\ref{stimegibbs}. The following theorem is the result about the short-time conservation of Gibbsianness for the 
time-evolved system $\mu'_{t,N}$.

\begin{thm}\label{thm2}
Suppose we have $\sqrt{2}C(F,g)\Bigl(1-e^{-nt}\Bigr)^{\frac{1}{2}}<1$, then the time-evolved 
system $\mu'_{t,N}$ will be Gibbsian and
$\g'_{1,t}$, the single-site kernel for $\mu'_{t,N}$, has the continuity estimate
\begin{equation}\begin{split}
\|\g'_{1,t}(\cdot|\nu'_1)-\g'_{1,t}(\cdot|\nu'_2)\|&\leq 2C(F,g)^2\left(1-e^{-nt}\right)^{
\frac{1}{2}}\|\nu'_1-\nu'_2\|.
\end{split}
\end{equation}
\end{thm}
Observe from Theorem \ref{thm2} that the time-evolved measure will be Gibbsian whenever either the initial 
interaction is weak or $t$ is small enough. The above result was only stated
in \cite{KULOP08} for the corresponding Curie-Weiss model. We present this case below.
For the Curie-Weiss rotator model the 
interaction for the initial system is given by 
\begin{equation}\label{cweiss}
\P(\nu)=F\bigl(\nu[\s^1_i],\cdots,\nu[\s^{n+1}_i]\bigr)=-\dfrac{\beta \sum_{j=1}^{n+1} \nu[\s^j_i]^2}{2},
\end{equation}
where $g_{j}(\s_i)=\s_i^j$ is the $j$th coordinate of the point $\s_i\in \S^{n}$ and 
 $l=n+1$. As a corollary to Theorem \ref{thm2} we have the following short-time Gibbsianness 
result for the Curie-Weiss rotator model under diffusive time evolution.

\begin{cor}\label{cor3}
Suppose we have $4\sqrt{2}\beta(n+1) e^\beta\Bigl(1-e^{-nt}\Bigr)^{\frac{1}{2}}<1$, then the time-evolved 
system $\mu'_{t,N}$ will be Gibbsian and
$\g'_{1,t}$, the single-site kernel for $\mu'_{t,N}$ has the continuity estimate
\begin{equation}\begin{split}
\|\g'_{1,t}(\cdot|\nu'_1)-\g'_{1,t}(\cdot|\nu'_2)\|&\leq 32\beta^2(n+1)^2 e^{2\beta}\left(1-e^{-nt}\right)^{
\frac{1}{2}}\|\nu'_1-\nu'_2\|.
\end{split}
\end{equation}

\end{cor}
Corollary \ref{cor3} is found in \cite{KULOP08} as Lemma 5.1.
This corollary is reminiscent of the result in Theorem 2.2 of \cite{KUL2}, where the Curie-Weiss model 
under independent spin-flip dynamics was studied. It is shown therein that if $\beta$ is small enough 
(weak initial interaction), 
then the time-evolved system will be Gibbsian forever, but if $\beta$ is large, then the time-evolved system 
will only be Gibbsian for short times. 

\subsubsection{ Conservation of Gibbsianness for $n-$vector mean-field models under fine local approximations}\label{example1}
Consider general $F$ and $g$ as above, and decompose $\S^n$ into countably many pairwise disjoint subsets (countries) as in Subsection 
\ref{fine} above.

Then with this notation it follows from Theorem \ref{generalint} that 

\begin{prop}\label{prop1}
If the quantity $\rho \, C(F,g)<1$, then the transformed system is Gibbsian and the single-site kernel $\g_1'$ satisfies the 
continuity 
estimate 
\begin{equation}\begin{split}
\|\g_1'(\cdot|\nu'_1)-\g_1'(\cdot|\nu'_2)\|&\leq \rho\, C(F,g)^2\rm{std}_{\a}\|\nu'_1-\nu'_2\|.
\end{split}
\end{equation}
\end{prop}
The above proposition can be found in Lemma 5.2 of \cite{KULOP08}. Thus the transformed system $\mu'_N$ will be Gibbsian if 
either the initial interaction $\P$ is weak or the local coarse-grainings (i.e. the $S_i$) are fine enough. In other words: 
If we have initial Gibbsian mean-field system with spins living on the sphere and we partition the sphere into countries, 
representing each country by a distinct point in $S'$, then the resultant transformed system will be Gibbsian if the countries are 
small enough. 

Let us mention in this context the result of Theorem 1.2 of \cite{HaggKu04} for the corresponding fuzzy Potts 
mean-field model. In that paper it was  shown 
that the transformed system will be Gibbsian at all temperatures whenever the sets of points 
contracted into single points by the fuzzy map have cardinality at most $2$. 

\subsection{Mean-field rotators in non-vanishing external magnetic field:
loss and recovery of Gibbsianness}

In this section  specialize to the quadratic mean-field rotator model on the circle, 
where we focus now on the interesting case $h\neq 0$. In fact, although we do 
not treat the simpler case $h=0$ here, one can in a very similar way prove 
loss of Gibbsianness, again just as in the lattice situation.\\  

We start with the measure 
\begin{equation}\label{mielke}\begin{split}
&\mu_{\b,h,N}(d \s_1,\dots, d\s_N)\cr
&=\frac{1}{Z_{\b,h,N}} \exp\Bigl ( 
N\b m(\s_1,\dots,\s_N)^2 + N \b h \cdot m(\s_1,\dots,\s_N)) 
\bigr) \prod_{i=1}^N \a(d\s_i)
\end{split}
\end{equation}
where 
\begin{equation*}\begin{split}
&m(\s_1,\dots,\s_N)=\frac{1}{N}\sum_{i=1}^n \s_i
\end{split}
\end{equation*}
is a vector-sum in $\R^2$ and $\a(d\s_i)$ is the equidistribution.
We take a time-evolution with the transition  kernels $p_t(\s_i,\eta_i)$ describing 
Brownian motion on the circle, as above.   

We are interested in the Gibbsian character of the time-evolved measures 
\begin{equation*}\begin{split}
&\mu_{\b,h,t,N}(d \eta_1,\dots, d\eta_N)=\int \mu_{\b,h,N}(d \s_1,\dots, d\s_N)\prod_{i=1}^N p_t(\s_i,d\eta_i)
\end{split}
\end{equation*}
in the sense of continuity of limiting conditional kernels, as described above. 
The virtue of mean-field models is that we can describe the limiting kernels explicitly. 
By this we mean a description in terms of a minimization problem of an explicit expression.  
This has been done in the general setup of site-wise independent transformations 
in \cite{KULOP08}. For the present case 
we get the following concrete results. 

\begin{prop} The limiting kernels $\g'_{1,\b,h,t}(d \eta_1| \l)$ of the time-evolved mean-field models $\mu_{\b,h,t,N}$ are 
given by the formula 

\begin{equation}\label{gabi}\begin{split}
&\g'_{1,\b,h,t}(d \eta_1| \l)= 
\frac{\int e^{\b \s_1 \cdot (m^*(\b,h,t,\l)+h)}p_t( \s_1,d \eta_1)\a(d \s_1)}{\int e^{\b \s_1 \cdot (m^*(\b,h,t,\l)+h)}\a(d\s_1)}
\end{split}
\end{equation}
for all choices of the (non-negative) parameters $\b,h,t$ and the conditioning $\l$ (in the probability measures on 
the circle), for which  the minimizer (in the closed unit disk)
\begin{equation*}\begin{split}
&m^*(\b,h,t,\l)=\text{argmin}\Bigl\{
m \mapsto  F(m; \b,h,t,\l) \Bigr\}
\end{split}
\end{equation*}
is unique with
\begin{equation}\label{foo}\begin{split}
& F(m; \b,h,t,\l)=\b \frac{|m|^2}{2}- \int\l(d\eta_1)\log \int e^{\b \bar \s\cdot ( m + h )} p_t(\eta_1,d\bar \s).
\end{split}
\end{equation}
\end{prop}

We do not give a proof of $\eqref{gabi}$ here (which can be deduced from the general result in \cite{KULOP08}), but 
we sketch a fast heuristics which explains what happens:  
Note first that $F(m; \b,h,t,\l)$ denotes the rate function (up to an additive constant) of the initial model,  
constrained to have an empirical distribution $\l$ in the transformed model.
Conditioning the empirical distribution of the transformed spins outside the 
site $1$ to $\l$ produces a quenched system involving the initial spins which acquires the 
magnetisation $m^*(\b,h,t,\l)$. The propagation of the corresponding distribution of $\s_1$ to $\eta_1$ 
with the kernel $p_t$ gives the desired conditional probability distribution 
$\g'_{1,\b,h,t}( d \eta_1 |\l)$. 

\subsubsection{Gibbsianness at large times}

Compare the rate function \eqref{foo} to the well-known rate-function of 
the initial model \eqref{mielke}  given by  
\begin{equation}\label{mielkefoo}\begin{split}
& F_0(m; \b,h)=\b \frac{|m|^2}{2}  - \log \int e^{\b \bar \s\cdot ( m + h )} \a(d\bar \s).
\end{split}
\end{equation}
The map $m\mapsto F_0(m; \b,h)$ has a unique minimizer $m^*(\b,h)$, if $h\neq 0$ is arbitrary, pointing in the direction of $h$. 

The input to understand the large time-behavior is the fact that the kernel 
$p_t(\eta_i,d \s_i)$ converges to the equidistribution $\a(d\s_i)$, uniformly in $\eta_i$.  

From this we see that, at fixed $\b,h$,  the functions $m\mapsto F(m; \b,h,t, \l)$ 
converge to $m\mapsto F_0(m; \b,h)$, uniformly in $\l$. 
The same holds for higher derivatives w.r.t. $m$. 
These statements imply that, for $t$ sufficiently large, for all choices of $\l$ there 
is only one minimizer $m^*(\b,h,t,\l)$. Looking at the linear appearance of the measure $\l$ 
in \eqref{foo}, we see that $m^*(\b,h,t,\l)$ changes continuously under a  variation of $\l$. 

By the form of \eqref{gabi} this implies Gibbsianness.

\subsubsection{Non-Gibbsianness at intermediate times}

To prove non-Gibbsianness at the parameter-triple 
$(\b,h,t)$ we use the formula \eqref{gabi} for the limiting kernels 
for those quadruples $(\b,h,t,\l)$ where they are well-defined and, for fixed $(\b,h,t)$ 
we show that there exists a $\l=\l^\text{spec}$ at which the limiting kernels 
are not continuous. 

To do so, it suffices to exhibit a one-parameter trajectory $\e\mapsto \l_\e$ which is continuous in the weak topology s.t. 
\begin{enumerate}
\item $F(m; \b,h,t,\l_{\e})$ has unique minimizers for $\e$ in a neighborhood 
of $\e^\text{spec}$,  \\  excluding $\e^\text{spec}$
\item $\lim_{\e \uparrow \e^\text{spec}}m^*(\b,h,t,\l_{\e})\neq \lim_{\e \downarrow \e^\text{spec}}m^*(\b,h,t,\l_{\e})$ 
\end{enumerate}

So far, the reasoning is general. Now, to create a phase transition in the constrained model, 
a suitable choice of $\l$ which is able to balance the influence of the external 
magnetic field $h$ has to be found. We choose conditionings of the type
\begin{equation}\label{lachgas}\begin{split}
&\l_\e=\frac{1}{2}\d_{e(\pi + \e)}+ \frac{1}{2}\d_{e(\pi - \e)}\cr
\end{split}
\end{equation}
where $e(\theta)$ denotes the vector on the circle corresponding to the angle $\theta$.
This conditioning mimicks the choice of conditionings 
on $\Z^2$ obtained by putting $e(\pi + \e)$ on the even sublattice and 
$e(\pi - \e)$ on the odd sublattice. 

\begin{prop}
Let $\b$ large enough,  and $h=\bar h e(0)\neq 0$ be given. Then there exists a time-interval for which there exists 
an $\e(\b,\bar h,t)$ such that $m\mapsto F(m; \b, h, t, \l_\e)$ has two different global minimizers of the form $m=u e(0)$, 
pointing in the direction or in the opposite direction of $h$.   
\end{prop}

We provide an explanation of this phenomenon. Let us look at the rate-function 
for magnetisation-values pointing in the direction of $h$, in the conditioning $\l_\e$ which reads 
\begin{equation*}\begin{split}
&F( u e(0); \b, h, t, \l_\e)=\b \frac{u^2}{2}\cr 
& - \frac{1}{2}\log \int e^{\b \cos \theta ( u +\bar h )} q_t (\theta - (\pi + \e))d \theta 
- \frac{1}{2}\log \int e^{ \b \cos \theta ( u +\bar h )} q_t (\theta - (\pi - \e))d \theta \cr 
\end{split}
\end{equation*}
with the diffusion kernel on the sphere written in angular coordinates $\theta$ as: 
\begin{equation*}\begin{split}
&q_t(\theta)= \frac{1}{2 \pi }+\frac{1}{\pi}\sum_{k=1}^\infty e^{- k^2 t}\cos (k \theta)
\end{split}
\end{equation*}
For fixed parameters $\b,\bar h$, we use the new magnetization variable 
$U=\b(u +\bar h)$ to rewrite 
\begin{equation*}\begin{split}
&F( u e(0); \b, h, t, \l_\e)- \frac{\b \bar h^2}{2}
=  \frac{U^2}{2 \b}- U \bar h - L(U; \e,t) 
\end{split}
\end{equation*}
where 
\begin{equation*}\begin{split}
&L(U; \e,t)=\frac{1}{2}\log \int e^{U \cos( \theta + (\pi + \e))} q_t (\theta )d \theta 
+\frac{1}{2}\log \int e^{U \cos (\theta + (\pi - \e))} q_t (\theta)d \theta \cr 
\end{split}
\end{equation*}
The second term on the l.h.s. is an unimportant constant. 
This choice of parameters is handy because we have separated the influence  of 
the parameters, and moreover, two of them are appearing only linearly 
in our 4-parameter family.  

Let us explain how a balance between $\e$ and $\bar h$ can be used to create 
a situation of a pair of different equal depth-minimizers, without going into the 
details of the analysis of the regions in parameter-space where this can be done. 

For this heuristic argument, let us fix the $\e$ first. 
We note that $U\mapsto L(U; \e,t)$ is convex, so $\frac{U^2}{2 \b}- L(U; \e,t)$  
has a chance to have more than one local minimum, for good choices of $\b,\e,t$. 
Having found such a situation, a tuning of the $\bar h$ will 
result in a tilting of the rate-function which can 
create a situation where this pair has an  equal depth in the full function $\frac{U^2}{2 \b}- U\bar h - L(U; \e,t)$. 
The mechanism described provides us with a curve in the space of $\e$ and 
$\bar h$ where the two minima have equal depth. Now, fixing a value 
of $\bar h$ and varying the $\eps$ across this curve, yields a jump 
of the global minimizer which implies non-Gibbsianness. 

In the pictures, showing the plot of  $U\mapsto G(U;\b^{-1},\bar h,\e,t)$  we see this mechanism at work.

\begin{figure}[htp]
\includegraphics[height=4.5cm]{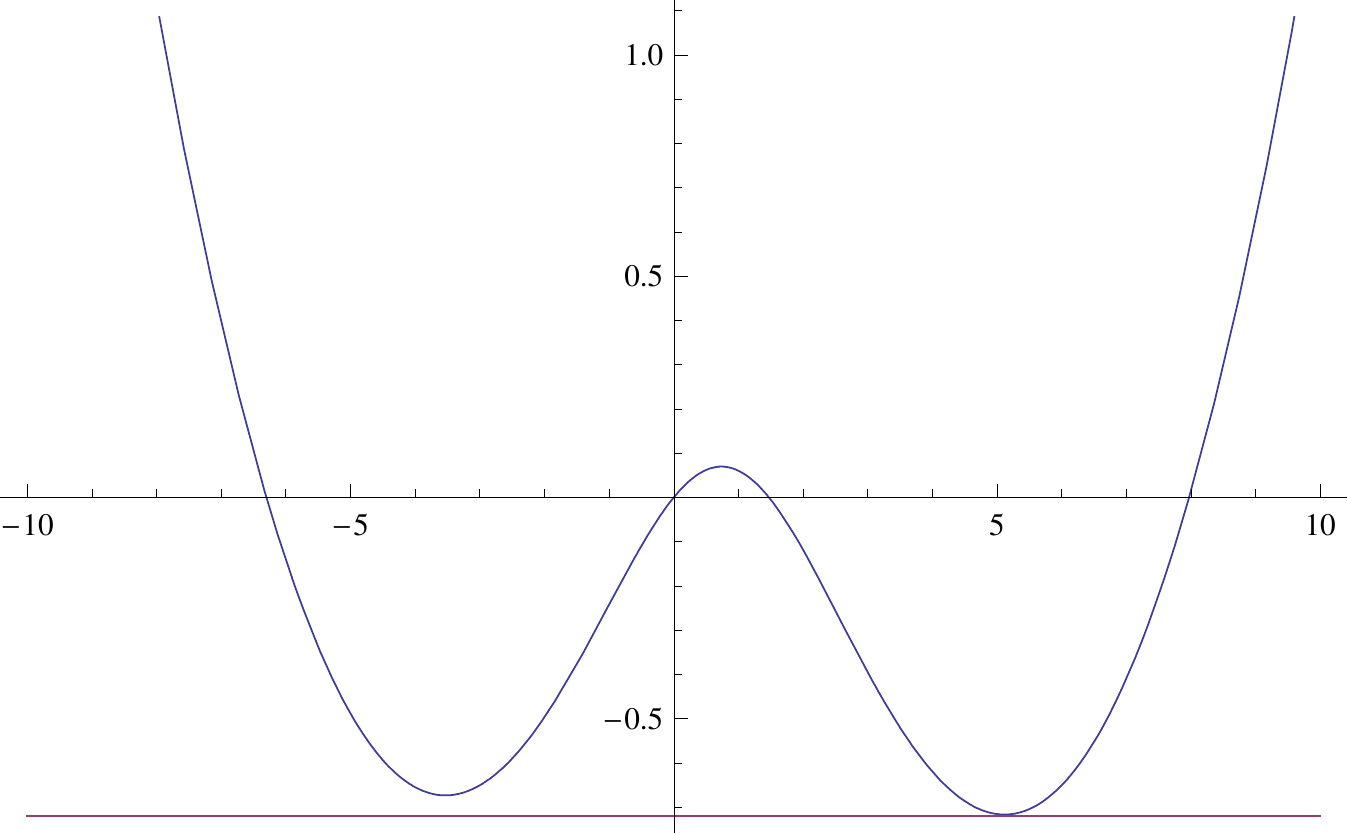}
\includegraphics[height=4.5cm]{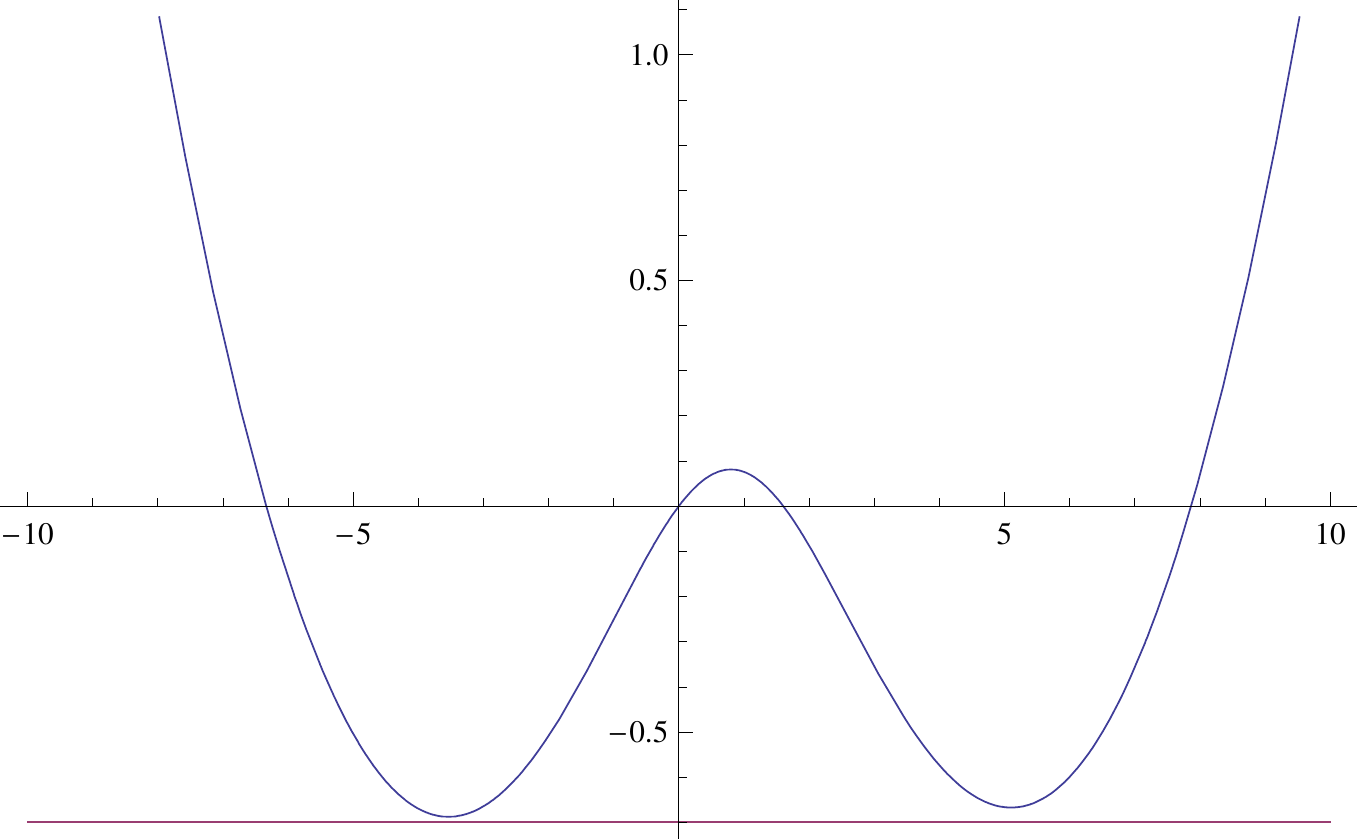}
$\phantom{123}$

 \caption{Left: $\b=5,\;\bar h=0.16,\; \e=0.4,\; t=1$, \;
Right: $\b=5,\;\bar h=0.16,\; \e=0.3,\; t=1$}
 \end{figure}
It is clear from the above diagram that for  $\b=5,\; \bar h=0.16,\;\text{and}\; t=1$ there is a choice of  $\e^*$
such that 
$F( u e(0); 5, .16, 1, \l_{\e*})$ has two global minimizers.
Numerically we find $\e^* \in (0.33481860,0.33481863)$. 
 Hence at such 
values for $\beta, h$ and $t$, 
the transformed system  
 will be non-Gibbsian.

\section*{Acknowledgements} We thank Roberto Fern\'andez and Frank Redig for many discussions. The possibility thereof was due to support from the NDNS mathematics research cluster.
  A.C.D. v.E. thanks Sacha Friedli and Bernardo de Lima for inviting him to a wonderful school 
in Ouro Preto and Maria Eulalia Vares for the invitation to write a review on the 
topics presented there.

\end{document}